\documentclass[preprints,article,accept,moreauthors,pdftex]{Definitions/mdpi} 

\firstpage{1} 
\pubvolume{1}
\issuenum{1}
\articlenumber{0}
\pubyear{2021}
\copyrightyear{2021}
\datereceived{} 
\dateaccepted{} 
\datepublished{} 
\hreflink{https://dx.doi.org/}
\pdfoutput=1

\usepackage{physics}
\usepackage{amssymb}
\usepackage{array}
\usepackage[list=false]{subcaption}
\usepackage[noprefix]{nomencl}
\usepackage{mathtools}

\usepackage{tikz-3dplot}
\usepackage{adjustbox}

\DeclarePairedDelimiter\ceil{\lceil}{\rceil}

\usetikzlibrary{calc, arrows.meta}
\usetikzlibrary{arrows}

\makenomenclature

\Title{Analysis of Carlemann Linearization of Lattice Boltzmann}

\TitleCitation{Title}
\Author{Wael Itani$^{1}$, Sauro Succi$^{2,3,4*}$}

\AuthorNames{Wael Itani, Antonio Mezzacapo, Sauro Succi}

\AuthorCitation{Itani, W.; Mezzacapo, A.; Succi, S.}

\address{%
$^{1}$ \quad Tandon School of Engineering, New York University, Brooklyn, 11201 New York, NY, USA \\
$^{2}$ \quad Center for Life Nano- \& Neuro-Science, Fondazione Istituto Italiano di Tecnologia (IIT), 00161 Rome, Italy \\
$^{3}$ \quad Istituto per le Applicazioni del Calcolo CNR, via dei Taurini 19, 00185, Rome, Italy \\
$^{4}$ \quad Institute for Applied Computational Science,
John A. Paulson School of Engineering and Applied Sciences, Harvard University, 02134 Boston, MA, USA}

\corres{Correspondence: sauro.succi@gmail.com}

\abstract{
We explore the Carlemann linearization of the collision term of the lattice Boltzmann formulation, as a first step towards formulating a quantum lattice Boltzmann algorithm.
Specifically, we deal with the case of a single, incompressible fluid with the Bhatnagar Gross and Krook equilibrium function. 
Under this assumption, the error in the velocities is proportional to the square of the Mach number. 
Then, we showcase the Carlemann linearization technique for the system under study. 
We compute an upper bound to the number of variables as a function of the order of the Carlemann linearization. 
We study both collision and streaming steps of the lattice Boltzmann formulation under Carlemann linearization. 
We analytically show why linearizing the collision step sacrifices the exactness of streaming in lattice Boltzmann, while also contributing to the blow 
up in the number of Carlemann variables in the classical algorithm. The error arising from Carlemann linearization
 has been shown analytically and numerically to improve exponentially with the Carlemann linearization order.
This bodes well for the development of a corresponding quantum computing algorithm based on the Lattice Boltzmann equation.
}

\keyword{Lattice Boltzmann; Carlemann; Linearization} 

\begin{document}
\makenomenclature

\maketitle

\nomenclature{$\vec{v}$}{continuum particle velocity}
\nomenclature{$\vec{c}_i$}{discrete velocity in the $i^{th}$ direction}
\nomenclature{$Q$}{number of discrete velocities number of modes at each lattice site, indexed by $i$}
\nomenclature{$D$}{the number of dimensions of the lattice}
\nomenclature{$x$}{the dimensions of the lattice, indexed by d, independent position vector variable}
\nomenclature{$N$}{number of Carlemann variables}
\nomenclature{$N_{x_d}$}{the number of sites across the $d^{th}$ dimension $x_d$ of the lattice}
\nomenclature{G}{the volume of the lattice in the units obtained by the product of the number of sites in each dimension $\Pi_{d=1}^D N_{x_d}$}
\nomenclature{$f_i$}{discrete density distribution weight}
\nomenclature{$\Omega$}{the collision operator defined by $\frac{d\vec{f}}{dt} = \Omega (\vec{f})$}
\nomenclature{$w_i$}{the weight of the $i^{th}$ discrete density}
\nomenclature{$O_c$}{truncation order in Carlemann linearization}
\nomenclature{$R$}{a measure of nonlinearity parametrizing the error bound of the Carlemann technique}
\nomenclature{$F_1$}{coefficient matrix of first-order terms in a quadratic ODE}
\nomenclature{$F_2$}{coefficient matrix of first-order terms in a quadratic ODE}
\nomenclature{$t$}{independent time variable}
\nomenclature{$\vec{u}$}{flow velocity}
\nomenclature{$\rho$}{local fluid density in lattice units}
\nomenclature{$c$}{lattice speed}
\nomenclature{$C$}{Carlemann linearization matrix}
\nomenclature{$\varepsilon$}{norm of the solution error}
\nomenclature{${f}_C$}{approximated solution of the system}
\nomenclature{$\Delta t$}{discrete timestep}
\nomenclature{$V$}{vector of Carlemann variables}
\nomenclature{$\vec{e}$}{lattice vectors}
\nomenclature{$Ma$}{Mach number}
\nomenclature{$U$}{unitary operator}
\nomenclature{$K$}{scaling factor of the logistic equation}
\nomenclature{$T$}{Total integration time}
\nomenclature{$p$}{order of the polynomial describing the driving function $\Omega$}
\section{Introduction}

Computational Fluid Dynamics (CFD) has been around for as long as 
computers exist, starting with von Neumann's program to simulate the weather
on the ENIAC machine (1950's) and even earlier, with 1922 Richardson's description
of "human" computers computing the weather by hand - he estimated that 64,000
human calculators, each calculating at a speed of 0.01 Flops/s, would be sufficient
to predict the weather in real time \cite{noauthor_weather_nodate}.
Leaving aside human calculators, electronic ones have made a 
spectacular ride till current days, from the few hundred Flops of ENIAC
to the current few hundred Petaflops of the Top 1 IBM-NVIDIA Summit computer.
Sixteen orders of magnitude in 70 years, close to a sustained Moore's law rate
(doubling every 1.5 years)!
Amazingly, CFD has been consistently on the forefront of such spectacular ride
and continues to do so to this day. 
However, when it comes with quantum computing, CFD does not appear to have captured
substantial attention to date.
In this paper we present a brief survey of current ongoing research work in this direction and a 
preparatory technique, known as Carlemann linearization, aimed at the development of a quantum computing
algorithm for the Lattice Boltzmann method for fluid flows. 

\subsection{Early Attempts for Quantum Simulation of Fluids}
The earliest attempts at quantum simulation of fluids have been based on the lattice gas or lattice Boltzmann algorithms. The first quantum lattice Boltzmann scheme dates back to the 1990s \cite{benzi_lattice_1992,succiLatticeBoltzmannEquation1993}. Around the turn of the millenium, Yepez demonstrated fluid dynamic simulations on a special-purpose quantum computer based on nuclear magnetic resonance (NMR) \cite{yepezQuantumComputationFluid1999}, using the quantum lattice gas algorithm \cite{vahalaQuantumLatticeGas2008,  yepezLatticeGasQuantumComputation1998}. Leading the trail, Yepez has also investigated Burger’s equation \cite{yepezEfficientQuantumAlgorithm2002,  yepezOpenQuantumSystem2006}, and entropic lattice Boltzmann models \cite{boghosianEntropicLatticeBoltzmann2001}. The latter has found its use in simulation of quantum fluid dynamics and other quantum systems \cite{vahalaQuantumLatticeGas2008}. 
We have recently seen a divergence towards citing Navier-Stokes as a future direction for papers discussing solving nonlinear differential equations on quantum computers, away from the early physically-motivated algorithms for fluid simulation, as the quantum lattice gas one mentioned above \cite{yepezEfficientQuantumAlgorithm2002,  yepezLatticeGasQuantumComputation1998,  yepezOpenQuantumSystem2006,  yepezQuantumComputationFluid1999}. However, the attempts to revive the physically-motivated algorithms beyond quantum systems, by \cite{steijlQuantumAlgorithmsFluid2020,mezzacapoQuantumSimulatorTransport2015,budinskiQuantumAlgorithmAdvection2021,lloydQuantumAlgorithmNonlinear2020} and others, are promising. The work of \cite{mezzacapoQuantumSimulatorTransport2015} stands out as it presents itself as a method not of quantum computation per se, but of quantum simulation. The latter leverages the correspondence between the Dirac and lattice Boltzmann equations. We, thus, find a compelling reason \cite{succi_lattice_2015} to explore lattice Boltzmann as the basis for quantum simulation of fluids, starting with its linearization, explored classically in this paper.
\subsection{Carlemann Linearization}

Carleman linearization appears to cast linearization of a function through Taylor series expansion into matrix form, suitable for use in defining state estimator of a non-linear system of known dynamics as part of the Koopman operator approach, in what is known as Carlemann-Koopman operator. The basic idea of Carlemann linearization is to introduce powers of the original variable as a variables in the system. The recurrence through which the new equations in the system is defined leads to an infinite dimensional system. The latter is prone to admitting solutions different than those of the original system \cite{steebLinearizationProcedureNonlinear1981}. The linearization is achieved at the cost of infinite-dimensions. Upon truncation of the resulting infinite-dimensional upper-triangular matrix \cite{foretsExplicitErrorBounds2017}, the accuracy of the approximation also suffers, deteriorating with time, making the truncated system most suitable for asymptotically stable stationary solutions \cite{steebLinearizationProcedureNonlinear1981a}. \cite{foretsExplicitErrorBounds2017}, through a power series approach, obtained the error bound, for a polynomial ODE reduced to a quadratic one, showing that it depends on the initial condition, and exponentially on time. They recommend discretizing the solution in time, and evaluating other basis functions. Their work has readily been extended by \cite{liuEfficientQuantumAlgorithm2020} for a quantum algorithm. Apart from discussing the complexity and error bounds of a quantum Carlemann algorithm, \cite{liuEfficientQuantumAlgorithm2020} presented the results for a classical Carlemann linearization of Burger's equation.

\section{Lattice Boltzmann}
Between the Navier-Stokes equations which model the flow at a continuum level, and molecular dynamics which treat the microscale, LB stands out, representing the fluid as an ensemble of particles at the mesoscale. It stems from the minimal discretization of the Boltzmann kinetic equation. The lattice Boltzmann formulation is readily extensible for a range of physics, from the quantum to the relativistic, giving the method a versatility about which books could be written \cite{succiLatticeBoltzmannEquation2018}.
\begin{equation}
\label{BolEq}
    \frac{d{f}}{dt} = \frac{\partial{{f}}}{\partial{t}}+\vec{v}\cdot \vec{\nabla} {f} = {\Omega}
\end{equation}
The Boltzmann Eq.~\ref{BolEq} describes the probability density $f$ of the fluid in the position-momentum space, driven by advection due to continuum particle velocity $\vec{v}$ and collision $\Omega$ across space spanned by $\vec{x}$, and time $t$.  To arrive at the lattice Boltzmann formulation, the probability density $f$ from the Boltzmann equation Eq.~(\ref{BolEq}) is discretized into $Q$ density distributions, each describing the fraction of fictitious particles: moving in a given $D$-dimensional lattice, with $\vec{v}$ restricted to speeds $\vec{c}_i$, $\vec{c}_i=c_i\vec{e}_i$, defined in the directions of the lattice vectors $\vec{e}_i$s. 

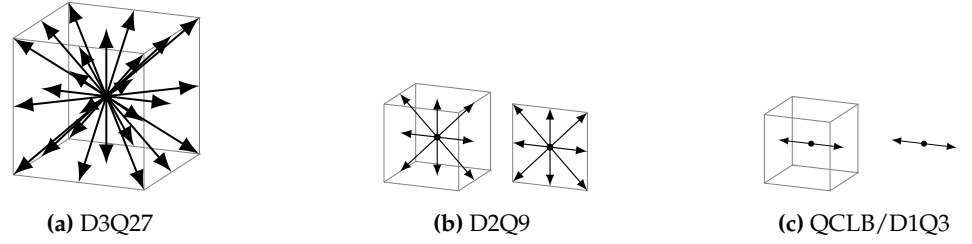
\begin{figure}[ht]
\centering
\captionsetup[subfigure]{justification=centering}

\subcaptionbox[D3Q27]{D3Q27\label{subfig:d3q27}}
[3cm]
{\resizebox{3cm}{!}{
\tdplotsetmaincoords{74}{113}
         \begin{tikzpicture}[tdplot_main_coords,
axis/.style={thick, ->, >=stealth'}]

\coordinate (O) at (0.5,0.5,0.5);

\def \a {1}       
\def \b {1}       
\def \c {1}       

 \foreach \u in {0,1,...,\a}
    \foreach \v in {0,1,...,\b}
      \foreach \w in {0,1,...,\c}
        \draw[very thin,gray] (\u,\v,0) -- (\u,\v,\w); 
 \foreach \u in {0,1,...,\a}
    \foreach \v in {0,1,...,\b}
      \foreach \w in {0,1,...,\c}
        \draw[very thin,gray] (\u,0,\w) -- (\u,\v,\w); 
 \foreach \u in {0,1,...,\a}
    \foreach \v in {0,1,...,\b}
      \foreach \w in {0,1,...,\c}
        \draw[very thin, gray] (0,\v,\w) -- (\u,\v,\w);

\draw plot [mark=*, mark size=1] coordinates{(O)};

 \foreach \u in {0,1,...,\a}
    \foreach \v in {0,1,...,\b}
      \foreach \w in {0,1,...,\c}
        \draw[thin,-latex,black](O) -- (\u,\v,\w);
        
    \foreach \v in {0,1,...,\b}
      \foreach \w in {0,1,...,\c}
        \draw[thin,-latex,black](O) -- (0.5,\v,\w);
 \foreach \u in {0,1,...,\a}
      \foreach \w in {0,1,...,\c}
        \draw[thin,-latex,black](O) -- (\u,0.5,\w);
 \foreach \u in {0,1,...,\a}
    \foreach \v in {0,1,...,\b}
        \draw[thin,-latex,black](O) -- (\u,\v,0.5);

 \foreach \u in {0,1,...,\a}
        \draw[thin,-latex,black](O) -- (\u,0.5,0.5);
    \foreach \v in {0,1,...,\b}
        \draw[thin,-latex,black](O) -- (0.5,\v,0.5); 
      \foreach \w in {0,1,...,\c}
        \draw[thin,-latex,black](O) -- (0.5,0.5,\w); 

\end{tikzpicture}}}
\hspace{0.1\textwidth}
\subcaptionbox[D2Q9]{D2Q9\label{subfig:d2q9}}
[3cm]
{\resizebox{3cm}{!}{
\tdplotsetmaincoords{74}{113}
         \begin{tikzpicture}[tdplot_main_coords,
axis/.style={thick, ->, >=stealth'}]

\coordinate (O) at (0.5,0.5,0.5);

\def \a {1}       
\def \b {1}       
\def \c {1}       

 \foreach \u in {0,1,...,\a}
    \foreach \v in {0,1,...,\b}
      \foreach \w in {0,1,...,\c}
        \draw[very thin,gray] (\u,\v,0) -- (\u,\v,\w); 
 \foreach \u in {0,1,...,\a}
    \foreach \v in {0,1,...,\b}
      \foreach \w in {0,1,...,\c}
        \draw[very thin,gray] (\u,0,\w) -- (\u,\v,\w); 
 \foreach \u in {0,1,...,\a}
    \foreach \v in {0,1,...,\b}
      \foreach \w in {0,1,...,\c}
        \draw[very thin, gray] (0,\v,\w) -- (\u,\v,\w);

\draw plot [mark=*, mark size=1] coordinates{(O)};

    \foreach \v in {0,1,...,\b}
      \foreach \w in {0,1,...,\c}
        \draw[thin,-latex,black](O) -- (0.5,\v,\w);

    \foreach \v in {0,1,...,\b}
        \draw[thin,-latex,black](O) -- (0.5,\v,0.5); 
      \foreach \w in {0,1,...,\c}
        \draw[thin,-latex,black](O) -- (0.5,0.5,\w); 

\end{tikzpicture}

\tdplotsetmaincoords{74}{113}
         \begin{tikzpicture}[tdplot_main_coords,
axis/.style={thick, ->, >=stealth'}]

\coordinate (O) at (0.5,0.5,0.5);

\def \a {1}       
\def \b {1}       
\def \c {1}       

 \foreach \v in {0,1,...,\b}
      \foreach \w in {0,1,...,\c}
        \draw[very thin,gray] (0.5,\v,0) -- (0.5,\v,\w); 
 \foreach \v in {0,1,...,\b}
      \foreach \w in {0,1,...,\c}
        \draw[very thin,gray] (0.5,0,\w) -- (0.5,\v,\w);

\draw plot [mark=*, mark size=1] coordinates{(O)};

    \foreach \v in {0,1,...,\b}
      \foreach \w in {0,1,...,\c}
        \draw[thin,-latex,black](O) -- (0.5,\v,\w);

    \foreach \v in {0,1,...,\b}
        \draw[thin,-latex,black](O) -- (0.5,\v,0.5); 
      \foreach \w in {0,1,...,\c}
        \draw[thin,-latex,black](O) -- (0.5,0.5,\w); 

\end{tikzpicture}}}
\hspace{0.1\textwidth}
\subcaptionbox[D1Q3]{QCLB/D1Q3\label{subfig:d1q3}}
[3cm]
{\resizebox{3cm}{!}{
\tdplotsetmaincoords{74}{113}
         \begin{tikzpicture}[tdplot_main_coords,
axis/.style={thick, ->, >=stealth'}]

\coordinate (O) at (0.5,0.5,0.5);

\def \a {1}       
\def \b {1}       
\def \c {1}       

 \foreach \u in {0,1,...,\a}
    \foreach \v in {0,1,...,\b}
      \foreach \w in {0,1,...,\c}
        \draw[very thin,gray] (\u,\v,0) -- (\u,\v,\w); 
 \foreach \u in {0,1,...,\a}
    \foreach \v in {0,1,...,\b}
      \foreach \w in {0,1,...,\c}
        \draw[very thin,gray] (\u,0,\w) -- (\u,\v,\w); 
 \foreach \u in {0,1,...,\a}
    \foreach \v in {0,1,...,\b}
      \foreach \w in {0,1,...,\c}
        \draw[very thin, gray] (0,\v,\w) -- (\u,\v,\w);

\draw plot [mark=*, mark size=1] coordinates{(O)};

    \foreach \v in {0,1,...,\b}
        \draw[thin,-latex,black](O) -- (0.5,\v,0.5); 
\end{tikzpicture}

\tdplotsetmaincoords{74}{113}
         \begin{tikzpicture}[tdplot_main_coords,
axis/.style={thick, ->, >=stealth'}]

\coordinate (O) at (0.5,0.5,0.5);

\def \a {1}       
\def \b {1}       
\def \c {1}       

 \foreach \u in {0,1,...,\a}
    \foreach \v in {0,1,...,\b}
      \foreach \w in {0,1,...,\c}
        \draw[very thin,white] (\u,\v,0) -- (\u,\v,\w); 
 \foreach \u in {0,1,...,\a}
    \foreach \v in {0,1,...,\b}
      \foreach \w in {0,1,...,\c}
        \draw[very thin,white] (\u,0,\w) -- (\u,\v,\w); 
 \foreach \u in {0,1,...,\a}
    \foreach \v in {0,1,...,\b}
      \foreach \w in {0,1,...,\c}
        \draw[very thin,white] (0,\v,\w) -- (\u,\v,\w);

\draw plot [mark=*, mark size=1] coordinates{(O)};

    \foreach \v in {0,1,...,\b}
        \draw[thin,-latex,black](O) -- (0.5,\v,0.5); 
\end{tikzpicture}}}
\caption{Different lattice configurations in three, two and one dimensions}
\label{fig:lattices}
\end{figure}

The discretized lattice Botlzmann equation takes the form
\begin{equation}
\label{lBolEq}
    \frac{1}{\Delta t}(f_i(\vec{x}+\vec{c}_i\Delta t, t + \Delta t) - f_i(\vec{x},t)) = - \frac{1}{\tau}(f_i(\vec{x},t)-f_i^{eq}(\vec{x},t))
\end{equation}
for the simple case of a single-phase fluid, which components are permeable to each other such that they can be considered well-mixed and homogeneous \cite{billLatticeBoltzmannMethod}. The discrete probability density is the fundamental variable of the lattice Boltzmann approach. It describes the probability of finding a fictitious fluid particle at a given point defined by the position vector $\vec{x}$, at a given instant of time $t$, with a particular speed \cite{randles_performance_2013}. Implicit in the discretization of the Boltzmann equation, is the discretization of the phase-space, involving discrete position and velocity values.
The lattice Boltzmann proceeds by updating the discrete probability densities at each cell in two steps: collision and advection/streaming. This is the collision step:
\begin{equation}
\label{colBolEq}
    f_i(\vec{x}, t + \Delta t) = f_i(\vec{x},t) -\frac{\Delta t}{\tau}(f_i(\vec{x},t)-f_i^{eq}(\vec{x},t))
\end{equation}
And the streaming step
\begin{equation}
\label{strBolEq}
    f_i(\vec{x},t + \Delta t) \xrightarrow{} f_i(\vec{x}+\vec{c}_i\Delta t, t + \Delta t)
\end{equation}

\begin{figure}[ht]
\centering
\captionsetup[subfigure]{justification=centering}
\subcaptionbox[Pre-Collision]{Pre-Collision\label{subfig:precollision}}
[3cm]
{\resizebox{3cm}{!}{
\tdplotsetmaincoords{0}{0}
         \begin{tikzpicture}[tdplot_main_coords,
axis/.style={thick, ->, >=stealth'}]

\coordinate (O) at (0.5,0.5,0.5);

 \foreach \u in {0,1,...,1}
        \draw[very thin,gray] (\u-0.5,0.5,0.5) -- (\u+0.5,0.5,0.5); 

 \foreach \v in {-0.5,0.5,...,1.5}
\draw plot [mark=*, mark size=1] coordinates{(\v,0.5,0.5)};

\draw[ultra thick,-latex,red] (1,0.5,0.5) -- (O);
\draw[ultra thin,-latex,blue](0,0.5,0.5) -- (O);
\end{tikzpicture}}}
\hspace{0.1\textwidth}
\subcaptionbox[Post-Collision]{Post-Collision\label{subfig:postcollision}}
[3cm]
{\resizebox{3cm}{!}{
\tdplotsetmaincoords{0}{0}
         \begin{tikzpicture}[tdplot_main_coords,
axis/.style={thick, ->, >=stealth'}]

\coordinate (O) at (0.5,0.5,0.5);

 \foreach \u in {0,1,...,1}
        \draw[very thin,gray] (\u-0.5,0.5,0.5) -- (\u+0.5,0.5,0.5); 

 \foreach \v in {-0.5,0.5,...,1.5}
\draw plot [mark=*, mark size=1] coordinates{(\v,0.5,0.5)};

\draw[thick,-latex,red] (O) -- (0,0.5,0.5);
\draw[thin,-latex,blue] (O) -- (1,0.5,0.5);
\end{tikzpicture}}}
\hspace{0.1\textwidth}
\subcaptionbox[Streaming]{Streaming\label{subfig:streaming}}
[3cm]
{\resizebox{3cm}{!}{
\tdplotsetmaincoords{0}{0}
         \begin{tikzpicture}[tdplot_main_coords,
axis/.style={thick, ->, >=stealth'}]

\coordinate (O) at (0.5,0.5,0.5);

 \foreach \u in {0,1,...,1}
        \draw[very thin,gray] (\u-0.5,0.5,0.5) -- (\u+0.5,0.5,0.5); 

 \foreach \v in {-0.5,0.5,...,1.5}
\draw plot [mark=*, mark size=1] coordinates{(\v,0.5,0.5)};

\draw[thick,-latex,red] (0,0.5,0.5) -- (-0.5,0.5,0.5);
\draw[thin,-latex,blue] (1,0.5,0.5) -- (1.5,0.5,0.5);
\end{tikzpicture}}}
\caption{An illustration of the D1Q3 lattice Boltzmann scheme}
\label{fig:d1q3scheme}
\end{figure}
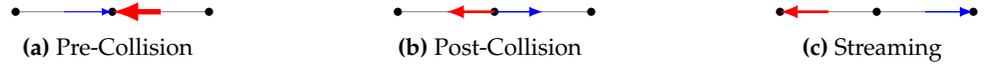

Collision is practically a relaxation towards equilibrium, a nonlinear operation dependent on terms local to each cell. On the other hand, streaming involves the transfer of the discrete densities to nearby cells, a nonlocal linear operation. The equilibrium distribution $f^{eq}_i$ is written as a function of $c_i=c$ the lattice speeds, the fluid density $\rho$, the lattice vectors $\vec{e}_i$, and the flow velocity $\vec{u}$.
\begin{equation}
    \rho = {\Sigma}_{i=1}^Q f_i
\end{equation}
flow velocity $\vec{u}$:
\begin{equation}
    \vec{u}=\frac{c}{\rho} {\Sigma}_{i=1}^Q f_i\vec{e}_i
\end{equation}
weights $w$ which sum to unity. A common model for the equilibrium function is:
\begin{equation}
    f_i^{eq}(\vec{x},t) = w_i\rho(1+(\frac{3}{c}\vec{e}_i\cdot\vec{u}+\frac{9}{2c^2}(\vec{e}_i\cdot \vec{u})^2-\frac{3}{2c^2}\vec{u}\cdot\vec{u}))
\end{equation}
Replacing the expression for the equilibrium expression for the incompressible case, we have:
\begin{equation}
\label{approxO}
\begin{split}
    (f_i(\vec{x}+\vec{c}_i\Delta t, t + \Delta t) =& (1-\frac{\Delta t}{\tau})f_i(\vec{x},t)) 
    \\&+ \frac{\Delta t w_i}{\tau}(1+3\vec{e}_i\cdot f_j\vec{e}_j+\frac{9}{2}(\vec{e}_i\cdot f_j\vec{e}_j)^2-\frac{3}{2}f_jf_k\vec{e}_j\cdot\vec{e}_k)
    \end{split}
\end{equation}
We note that the lattice vectors $\vec{e}_i$, the column vectors of $e$, do not correspond to unit vectors. For example, in D1Q3, we have:
\begin{equation}
e = 
    \begin{pmatrix}
-1 & 0 & 0\\
0 & 0 & 1 \\
\end{pmatrix}
\end{equation}
thereby:
\begin{equation}
\Omega_i =
\begin{cases}
-\frac{dt}{2\tau}
(f1 + f3 - (f1 - f3)^2 - \frac{1}{3})\\
-\frac{dt}{\tau}(f2 + (f1 - f3)^2 -\frac{2}{3})\\
-\frac{dt}{2\tau}(f1 + f3 - (f1 - f3)^2 - \frac{1}{3})
\end{cases}
\end{equation}
It is well-known that Eq. \ref{approxO} is inconsistent with the incompressible assumption, $\rho = 1$, and density fluctuations around unity are expected to arise during evolution \cite{he_lattice_1997}. Neglecting these fluctuations amounts to an error in the velocity components $||\vec{u}||$ proportional to the square of the Mach number $O(Ma^2)$. This is due to the fact that this standard lattice Boltzmann formulation recovers the compressible Navier-Stokes upon expansion. More sophisticated formulations have been developed for incompressible and nearly incompressible flows \cite{lallemand_lattice_2021}.
The particles with random motion, are restricted to the lattice nodes with microscopic velocities $c_i$ defined over lattice directions, allowing us to model the collision of particles and their streaming in seperate, uncoupled steps. The latter forces the nonlinearity of fluid flow, captured in the collision term, to be local, whereas the non-local streaming terms remain linear. Moreover, streaming is exact.

\subsection{Nonlinearity Ratio}
We define the nonlinearity ratio $R$ as a measure of how much Eq.~(\ref{approxO}) deviates from a complete linear behavior. It is defined as 
\begin{equation}
    R = ||f(t=0)|| \frac{||F_2||}{|\textrm{Real}(\lambda_{MAX}(F_1))|},
\end{equation}
where $F_1$ is the matrix of linear coefficients of Eq.~(\ref{approxO}) and $F_2$ is the matrix of the second order terms. Here we have considered the supremum norm of a matrix  $||A||$ or a vector $||\vec{x}||$ as 
\begin{equation}
||A|| = \max_{ij} |A_{ij}|, \; ||\vec{x}|| = \max_{i} |x_i|
\end{equation}
and $\lambda_{MAX}(F_1)$ as the largest eigenvalue in modulo of $F_1$. Note that for a linear system $||F_2|| = 0$ and one has no nonlinearity. We note that this make $R$ "qualitatively" similar to the Reynolds number which is a ratio of nonlinear convective forces to linear viscous forces.

\section{Carlemann linearization for lattice Boltzmann}

The basic tenant of Carlemann linearization is a change of variables, done such that the variables of the original nonlinear system, $\vec{f}(t)$, are replaced by a larger set of variables $\vec{V}(t)$. The original variables form a subset of the larger set of Carlemann variables. The additional variables are monomials of order $O_c$ of the original $f_i$. 

\begin{figure}[ht]
    \centering
    \begin{equation*}
    \begin{pmatrix}
          &        & | & f_1^2f_3
      \\    & f_1f_3 & | & f_1^3
      \\f_1 & f_1^2  & | & f_1^2f_2
      \\    & f_1f_2 & | & f_1f_2^2
      \\f_2 & f_2^2  & | & f_2^3
      \\    & f_2f_3 & | & f_2^2f_3
      \\f_3 & f_3^2  & | & f_2f_3^2
      \\    &        & | & f_3^3
      \\    &        & | & f_3^2f_1
    \end{pmatrix} \longrightarrow
    \begin{pmatrix}
      V_1 \\ V_2 \\ V_3 \\ V_4 \\ V_5 \\ V_6 \\ V_7 \\ V_8 \\ V_9
    \end{pmatrix}
    \end{equation*}
    \caption{Illustration of the how the Carlemann variables are introduced to linearize terms up to a chosen truncation order, in this case 2. In the case of lattice Boltzmann, the terms of the same order depend on terms at most one order higher (Each column depends on all columns that precede it and the one right after it).}
    \label{fig:my_label}
\end{figure}

An example of additional variables of second order is shown in Table \ref{tab:2ndCarl} for a D1Q3 lattice, taking into account the model in Eq.~(\ref{approxO}). 

\begin{figure}[ht]
\centering
{\resizebox{3cm}{!}{
\tdplotsetmaincoords{0}{0}
         \begin{tikzpicture}[tdplot_main_coords,
axis/.style={thick, ->, >=stealth'}]

\coordinate (O) at (0.5,0.5,0.5);
\draw plot [mark=*, mark size=1] coordinates{(O)} node [below] {$f_2$};

\draw[thick,-latex,red] (O) -- (0,0.5,0.5) node [below left] {$f_1$};
\draw[thin,-latex,blue] (O) -- (1,0.5,0.5) node [below right] {$f_3$};
\end{tikzpicture}}}
\caption{Numbering convention used for the discrete densities of the D1Q3 lattice Boltzmann scheme}
\label{fig:numbering}
\end{figure}
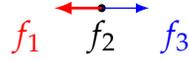

The dynamics of the extended system are then derived from the original system of a single phase, homogeneous, fluid, with terms beyond a chosen order $O_c$ dropped. 

\begin{equation}
\label{Eq::density_time_derivative}
    \frac{\partial f_i(t)}{\partial t} = \Omega_i(\vec{f}(t))
\end{equation}
which is linearized into the total derivative of the Carlemann variables vector $\vec{V}(t)$ equating a constant coefficient matrix, the Carlemann linearization matrix $C$, multiplying the Carlemann variables vector:
\begin{equation}
\label{Carl_Equation}
    \frac{\partial \vec{V}(t)}{\partial t} = C\vec{V}(t)
\end{equation}
where the Carlemann linearization matrix $C$ is obtained by deriving the following system of equations:
\begin{equation}
\label{Eq::Carlemann_var_def}
    \frac{\partial V_n}{\partial t}  = \Sigma_{i=1}^Q\frac{\partial V_n(t)}{\partial f_i}\Omega_i(f(t)) 
\end{equation}
and identifying the variables $V_i$ on the right hand side of Eq.~(\ref{Eq::Carlemann_var_def}).

\begin{table}[ht]
  \centering
    \caption{Example of D1Q3 expanded to a second order truncation in Carlemann linearization, with $N = 9$ Carlemann variables. Note that the dummy variable $V_1=1$ is defined to simplify the form of Eq.~(\ref{Carl_Equation}).}
  \begin{tabular}{cc}
    Carlemann Variables & Lattice Variables \\
    \hline
    $V_1$ & $1$     \\
    $V_2$ & $f_1^2$ \\
    $V_3$ & $f_3^2$ \\
    $V_4$ & $f_1f_2$ \\
    $V_5$ & $f_1f_3$ \\
    $V_6$ & $f_2f_3$ \\
    $V_7$ & $f_1$ \\
    $V_8$ & $f_2$ \\
    $V_9$ & $f_3$ \\
  \end{tabular}
  \label{tab:2ndCarl}
\end{table}

\subsection{Number of Variables}
An upper bound for the number of Carlemann variables for a desired order $O_c$, when considering $N$ original lattice variables, is

\begin{equation}
N =     \frac{(O_c+(Q+1)-1)!}{(O_c)!((Q+1)-1)!} 
    = \begin{pmatrix}    O_c + Q \\ Q +1    \end{pmatrix}
\end{equation}

This is an upper bound to the number of Carlemann variables used in the linearized system because in practice some terms of order $O_c$ do not appear in the linearized equations. For example, from Table~\ref{tab:2ndCarl} $f_2^2$, the second order term for the rest particles density,  does not appear in the D1Q3 formulation. In general, the exact number of variables is given specifying a lattice structure and an equilibrium function. 

\subsection{Carlemann Linearization of Collision Step}
The collision operator Eq.~`\ref{colBolEq} can be expressed as a function of the Carlemann variables $V_i$s. To do that, we go back to Eq.~\ref{BolEq}, and consider only . The total derivative of the discrete densities is described by the nonlinear collision operator on the right-hand side.

\begin{figure}[!ht]
\centering
{\resizebox{15cm}{!}{\includegraphics{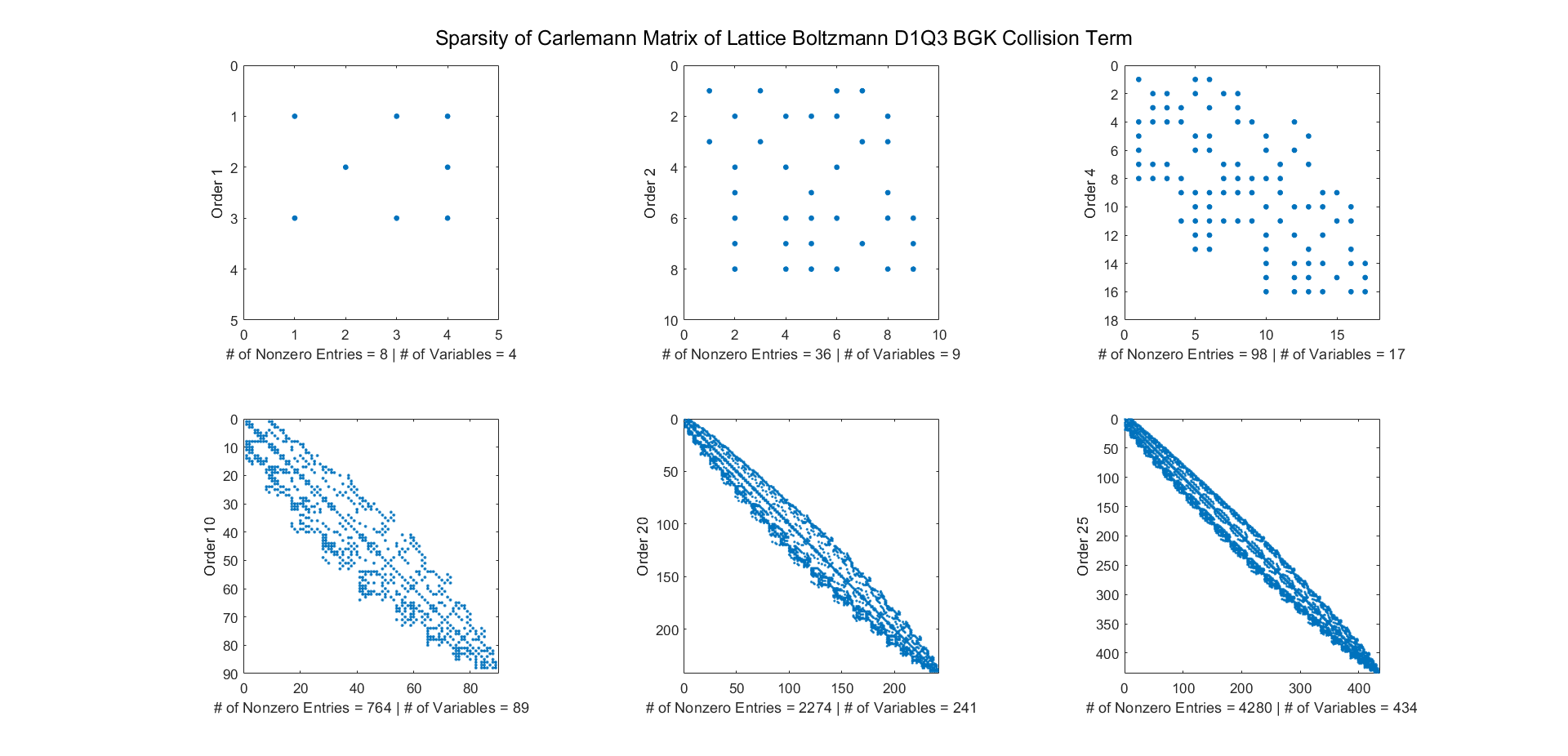}}}
\caption{Visualization of the sparsity of the Carlemann matrix for the collision term at various orders}
\end{figure}

This assumes that the obtained solution for the discrete density distribution at each site is streamed exactly, and the nonlinear terms recalculated to evolve the system. While this allows us to isolate the error from the linearization of the collision step, it is undesirable in practice. On another note, the linearization of the collision step allows one to explore other discretization scheme to arrive at the LB formulation from the Boltzmann equation Eq. \ref{BolEq}. In particular, we conjencture that an implicit scheme for the time discretization would improve the error bounds at the cost of the sparsity of the resulting matrix.
\subsection{Carlemann Linearization of Streaming Step}
Instead, we need to consider Eq. \ref{BolEq}. Another limitation of considering only a traditional one-dimensional model, such as Burger's is that the problem of streaming the coupled terms of the Carlemann linearization is avoided \cite{liuEfficientQuantumAlgorithm2020}. The forward and backward directions of the velocity are accounted for in one velocity variable with its negative and positive values, and derivatives of the velocity are discretized in terms of that one variable as well. 

We have thus far considered linearization of the collision step solely. However, for a self-contained quantum fluid simulation algorithm, Carlemann streaming must be considered, which leads us to a modified Boltzmann equation for the Carlemann variables:
\begin{equation}
   \frac{dV(\vec{x},t)}{dt} = \frac{\partial V(\vec{x},t)}{\partial t} + \vec{c}_i\cdot{\Sigma}_{i=1}^Q\frac{\partial V(\vec{x},t)}{\partial x_i} = CV
\end{equation} 
If the partial derivatives are left undiscretized, we have to include additional variables to account for the new terms appearing, contributing to the blowup of variables, and subjecting a description of streaming to truncation. On the other hand, following the same discretization scheme typical of LB, in similarity to the derivation of Eq. \ref{lBolEq}, we arrive at the a modified LB equation where the coefficients of the partial derivatives of the original variables $\vec{f}$ now show dependence on the variables themselves after replacing the expressions of the Carlemann variables $\vec{V}$ and computing their partial derivatives. Eq. \ref{ExNonLB} shows an example:
\begin{equation}
\label{ExNonLB}
       \frac{\partial V_n(\vec{x},t)}{\partial f_j (\vec{x},t)}\frac{\partial f_j (\vec{x},t)}{\partial t} + \vec{c}_i\cdot(\frac{\partial V_n(\vec{x},t)}{\partial f_j (\vec{x},t)}\frac{\partial f_j (\vec{x},t)}{\partial x_i}) = C_nV
\end{equation}
For D1Q3 second order linearization, with $n=4$, we have $V_4 = f_1f_2$:
\begin{equation}
\begin{split}
&f_2(\vec{x},t)(f_1(\vec{x}+\Delta\vec{x}_1,t)-f_1(\vec{x},t))
+f_1(\vec{x},t)(f_2(\vec{x}+\Delta\vec{x}_2,t)-f_2(\vec{x},t)) = 0
\\&f_2(\vec{x},t)f_1(\vec{x}+\Delta\vec{x}_1,t)-f_2(\vec{x},t)f_1(\vec{x},t)+f_1(\vec{x},t)f_2(\vec{x}+\Delta\vec{x}_2,t)-f_1(\vec{x},t)f_2(\vec{x},t) = 0
\\&f_2(\vec{x},t)f_1(\vec{x}+\Delta\vec{x}_1,t)-2V_4(\vec{x},t)
+f_1(\vec{x},t)f_2(\vec{x}+\Delta\vec{x}_2,t)= 0
\end{split}
\end{equation}
where we see new terms unaccounted for in the Carlemann variables appear combining both nonlocality, and nonlinearity.

Another way to explain the above is that the discrete densities of the particles are weighted by their contribution to the Carlemann variables, the new variables of the system $\vec{V}$, when streamed. When discretizing the  equation describing the evolution of the Carlemann variables, a coupling between terms at different locations appears due to the different partial derivatives appearing in the term. Therefore, classical Carlemann linearization of the lattice Boltzmann formulation exchanges the local nonlinearity of the collision step, with nonlocal linearity of the streaming step to which the linearized collision term is coupled. We note that additional $V$ variables must introduced for the nonlocal coupled terms, i.e. $f_1(\vec{x}_1+\Delta \vec{x}_1,t)f_2(\vec{x}_2,t)$ to keep the system linear. This is indeed used for the Burger's equation in previous literature \cite{liuEfficientQuantumAlgorithm2020}. However, this further exacerbates the blowup in variable count for the classical Carlemann scheme.

This leads to the fact that streaming is also described by an infinite differential system that must also be truncated. That is, the exactness of streaming, a major advantage of the lattice Boltzmann method, is lost. On a classical computer, to study the collision step, it is possible to recalculate the nonlinear terms to achieve exact streaming. Remarkably, while Carlemann linearization slashes out the exact streaming advantage of lattice Boltzmann, we are able to retrieve it by going into the quantum paradigm \citet{itani_quantum_2021}.

\subsection{Error Bound}
In Eq. \ref{approxO}, we see that LB fits into a generalized quadratic ODE considered by \cite{foretsExplicitErrorBounds2017} and \cite{liuEfficientQuantumAlgorithm2020}, and it is a multi-population extension of the logistic equation suggested by \cite{liuEfficientQuantumAlgorithm2020} for treatment.

For a given system of differential equations in terms of a set of variables represented by a vector $\vec{f}$, of which LB is an example, we denote the solution of the exact system as $\vec{f}$, and that of the Carlemann linearized system as $\vec{f}_C$. We use the same definition of the error $\varepsilon$ from \cite{foretsExplicitErrorBounds2017,liuEfficientQuantumAlgorithm2020}, in terms of the max supremum over time of the vector difference of exact and approximate solution normalized by their respective supremum norms:
\begin{equation}
\varepsilon(t) = \norm{\frac{\vec{f}(t)}{\norm{\vec{f}(t)}}-\frac{\vec{f}_C(t)}{\norm{\vec{f}_C(t)}}}
\end{equation}
As we integrate the system of equations, we define the maximum error made as 
\begin{equation}
    \varepsilon_{\max} = \max_{t\in T} \varepsilon(t)
    \end{equation}
When using Carlemann linearization and the Euler time discretization as in Eq.~(\ref{lBolEq}), the require \cite{liuEfficientQuantumAlgorithm2020} the minimum number of Carlemann variables $N$ will be a function of $\varepsilon_{\max}$ as  be:
\begin{equation}
\label{Neq}
        N = \ceil {-\frac{log_2(R)}{log_2(2(1+\frac{1}{\varepsilon_{\max}}))}}
\end{equation}
and the largest time-step $\Delta t$:
\begin{equation}
\label{DT}
\Delta t = \frac{1}{N\norm{F_1}} = (\ceil {-\frac{log_2(R)}{log_2(2(1+\frac{1}{\varepsilon}))}})\norm{F_1})^{-1}
\end{equation}

given that all the eigenvalues of $F_1$ are negative real, and $R<1$. 

The above formulas are derived for a single variable quadratic ODE system, $\vec{f} \in \mathcal{R}^1$, for which $Q=1$ and $N = O_c + 1$. Furthermore, in \cite{liuEfficientQuantumAlgorithm2020}, it is shown that the dependence of $N$ and $\Delta t$ with the error $\varepsilon_{\max}$ is of the form Eqs.(\ref{Neq}) and (\ref{DT}) even when $R>1$ for the Burgers equation. Note that for the LB problem the bounds of Eq.~(\ref{lBolEq}), Eq.~(\ref{Neq}) and Eq.~(\ref{DT}) are not valid as the quadratic system is always multivariate - $Q>1$ - and $R>1$ for typical choices of the equilibrium function, as for a single-phase fluid in Eq.~(\ref{approxO}).

Now we prove that the leading order in the error improves exponentially with the Carlemann order, for 1,2 and 3D systems.

Let $(r)$ denote a Carlemann variable $V_i$ of $r^{th}$ order, $V_i^{(r)}$, such that $C_{ij}^{(r)(k)}$ describes the matrix coefficient describing the contribution of $V_j^{(k)}$ to $V_i^{(r)}$. We have:
\begin{equation}
\begin{split}
    V_i^{(1)}(\vec{x},t) =&
    V_i^{(1)}(\vec{x},t-\Delta t)
    +\Sigma_j C_{ij}^{(1)(1)}V_j^{(1)}(\vec{x},t-\Delta t)
    \\&+ \Sigma_k C_{ik}^{(1)(2)}V_k^{(2)}(\vec{x},t-\Delta t)
\end{split}
\end{equation}
and:
\begin{equation}
\begin{split}
\label{Eq::secondordererror}
    V_i^{(2)}(\vec{x},t)=& V_i^{(2)}(\vec{x},t-\Delta t) 
    + \Sigma_jC_{ij}^{(2)(1)}V_j^{(1)}(\vec{x}+\Delta \vec{x}_j,t-\Delta t)
    \\&+\Sigma_kC_{ik}^{(2)(2)}V_k^{(2)}(\vec{x}+\Delta \vec{x}_k,t-\Delta t)
    \\&+\Sigma_lC_{il}^{(2)(3)}V_l^{(3)}(\vec{x}+\Delta \vec{x}_l,t-\Delta t)
    \\=& E_i^{(3)}(\vec{x},t-\Delta t) + V_i^{(2)}(\vec{x},t-\Delta t)
\end{split}
\end{equation}
such that when streaming is considered:
\begin{equation}
\begin{split}
\label{Eq::firstordererror}
    V_i^{(1)}(\vec{x},t) =& V_i^{(1)}(\vec{x},t-\Delta t)
    \\&+\Sigma_j C_{ij}^{(1)(1)}V_j^{(1)}(\vec{x}+\Delta\vec{x}_j,t-\Delta t)
    \\&+ \Sigma_k C_{ik}^{(1)(2)}V_k^{(2)}(\vec{x}+\Delta\vec{x}_k,t-\Delta t)
    \end{split}
\end{equation}
Replacing Eq.~\ref{Eq::secondordererror} into Eq.~\ref{Eq::firstordererror}, we have:
\begin{equation}
\begin{split}
    V_i^{(1)}(\vec{x},t) =& V_i^{(1)}(\vec{x},t-\Delta t)
    +\Sigma_j C_{ij}^{(1)(1)}V_j^{(1)}(\vec{x}+\Delta\vec{x}_j,t-\Delta t)
    \\&+ \Sigma_k C_{ik}^{(1)(2)}(E_k^{(3)}(\vec{x}+\Delta\vec{x}_k,t-2\Delta t) + V_k^{(2)}(\vec{x}+\Delta\vec{x}_k,t-2\Delta t))
    \end{split}
\end{equation}
For the collision problem considered in linearizing the collision term classically, $\Delta x_i$ = 0, such that one is able to verify:
\begin{equation}
\label{Eq:zeroerrorcond}
    \Sigma_jC_{ij}^{(1)(2)}E_j^{(2)}(\vec{x},t) = 0 
\end{equation}
for D1Q3, D2Q9 and D3Q27 "full"lattices for which the expressions become linear in Carlemann variables up to second order.  Eq.~\ref{Eq::firstordererror} then reduces to:
\begin{equation}
\begin{split}
    V_i^{(1)}(\vec{x},t) =& 
    V_i^{(1)}(\vec{x},t-\Delta t)
    +\Sigma_j C_{ij}^{(1)(1)}V_j^{(1)}(\vec{x}+\Delta\vec{x}_j,t-\Delta t)
    \\&+ \Sigma_k C_{ik}^{(1)(2)}(V_k^{(2)}(\vec{x}+\Delta\vec{x}_k,t-2\Delta t))
    \end{split}
\end{equation}
Dropping the $E$ term, and with further replacements similar to above, one can see that the first order terms depend only on the initial conditions of the second order terms in the domain, not only neighbouring cells, and no third order terms are needed. This is to say that turning off streaming resolves the nonlinearity of the problem, as expected. 

With streaming, a simple Taylor expansion of Eq.~(\ref{Eq::firstordererror}) shows that Carlemann linearization of order yields a solution with error of the order:
\begin{equation}
    \varepsilon_{max} = O(\Delta t \Delta x^{O_c})
\end{equation}.
If we initialize the flow to be uniform, the inclusion of streaming does not introduce errors as above, as Eq.~\ref{Eq:zeroerrorcond} still holds for the initial conditions at the cells are identical, and so is the collision step, such that $E(\vec{x},t)=E(\vec{x}+\Delta \vec{x}_1,t)=\dots=E(\vec{x}+\Delta \vec{x}_Q,t)$. At the boundaries, this still holds if they were periodic, but the latter amounts to the trivial case where the kinetic energy of the flow relaxes to zero. Periodic boundaries refer to a fully periodic domain, in all its dimensions, i.e. triply periodic in $3D$, which is typically useful for fundamental studies of homogeneous turbulence. 

\begin{figure}[ht]
\centering
\captionsetup[subfigure]{justification=centering}
\subcaptionbox{Uniform initial flow field\label{subfig:error_uniform}}
[3cm]
{\resizebox{3cm}{!}{
\tdplotsetmaincoords{0}{0}
         \begin{tikzpicture}[tdplot_main_coords,
axis/.style={thick, ->, >=stealth'}]

\coordinate (O) at (0.5,0.5,0.5);

\draw[very thin,gray]    (-1,0,0.5) -- (-1,1,0.5);
 \foreach \u in {0,1,...,2}
        {\foreach \v in {-0.5,0.5,...,1.5}
            {\draw[thick,-latex,red] (\u-0.5,0.5,0.5) -- (\u,0.5,0.5);
            \draw[thin,-latex,blue]  (\u-0.5,0.5,0.5) -- (\u-1,0.5,0.5);
            \draw[very thin,gray]    (\u,0,0.5) -- (\u,1,0.5);
            }}
\foreach \v in {-0.5,0.5,...,1.5}
    \draw plot [mark=*, mark size=1] coordinates{(\v,0.5,0.5)};

\end{tikzpicture}}}
\hspace{0.1\textwidth}
\subcaptionbox{Periodic boundary conditions\label{subfig:error_periodic}}
[3cm]
{\resizebox{3cm}{!}{
\tdplotsetmaincoords{0}{0}
         \begin{tikzpicture}[tdplot_main_coords,
axis/.style={thick, ->, >=stealth'}]

\coordinate (O) at (0.5,0.5,0.5);

\draw[very thin,gray]    (-1,0,0.5) -- (-1,1,0.5);
 \foreach \u in {0,1,...,2}
        {\foreach \v in {-0.5,0.5,...,1.5}
            {\draw[thick,-latex,red] (\u-1,0.5,0.5) -- (\u-0.5,0.5,0.5);
            \draw[thin,-latex,blue]  (\u,0.5,0.5) -- (\u-0.5,0.5,0.5);
            \draw[very thin,gray]    (\u,0,0.5) -- (\u,1,0.5);
            }}
\foreach \v in {-0.5,0.5,...,1.5}
    \draw plot [mark=*, mark size=1] coordinates{(\v,0.5,0.5)};

\end{tikzpicture}}}
\hspace{0.1\textwidth}
\subcaptionbox{Non-periodic boundary conditions\label{subfig:error_nonperiodic}}
[3cm]
{\resizebox{3cm}{!}{
\tdplotsetmaincoords{0}{0}
         \begin{tikzpicture}[tdplot_main_coords,
axis/.style={thick, ->, >=stealth'}]

\coordinate (O) at (0.5,0.5,0.5);

\draw[very thin,gray]    (-1,0,0.5) -- (-1,1,0.5);
 \foreach \u in {0,1,...,2}
        {\foreach \v in {-0.5,0.5,...,1.5}
            {\draw[very thin,gray]    (\u,0,0.5) -- (\u,1,0.5);
            }}
            
\draw[ very thick,-latex,red] (-1,0.5,0.5) -- (-0.5,0.5,0.5);
\draw[very thin,-latex,blue]  (0,0.5,0.5) -- (-0.5,0.5,0.5);

\draw[thick,-latex,red] (0,0.5,0.5) -- (0.5,0.5,0.5);
\draw[thin,-latex,blue]  (1,0.5,0.5) -- (0.5,0.5,0.5);

\draw[ultra thick,-latex,blue]  (2,0.5,0.5) -- (1.5,0.5,0.5);

\foreach \v in {-0.5,0.5,...,1.5}
    \draw plot [mark=*, mark size=1] coordinates{(\v,0.5,0.5)};

\end{tikzpicture}}}
\caption{Illustration using D1Q2 to show why the uniform initial flow with periodic boundaries coincides with the error-free case of collision without streaming while errors form at the boundaries when non-periodic boundary conditions are applied. We see that with the same initial conditions and periodic boundary, local and neighboring information of discrete densities are interchangeable}
\label{fig:errorillustration}
\end{figure}
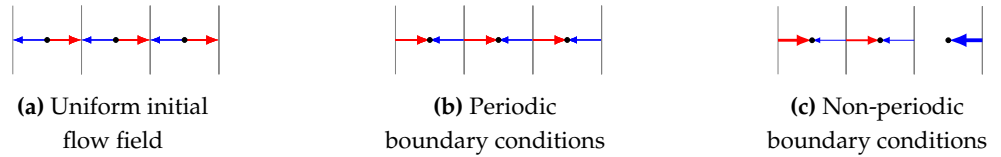

If (one of) boundaries are not periodic, the error is first generated in the collision step at the boundary, and propagated to the interior of the domain. With each time cycle, the error grows, and it propagates further inside, such that we may speak of a numerical error boundary layer. For example, in a pipe with periodic flow, where the error is first generated at the walls of the pipe, rather than the periodic inlet and outlet.

\begin{figure}[ht]
\centering
\captionsetup[subfigure]{justification=centering}
\subcaptionbox{Initial condition, at $t=0$\label{subfig:errorinitial}}
[3cm]
{\resizebox{3cm}{!}{\begin{tikzpicture}
\draw[style=help lines,dashed] (0,0) grid[step=1] (5,5);
\end{tikzpicture}}}
\hspace{0.1\textwidth}
\subcaptionbox{After one timestep, at $t=\Delta t$\label{subfig:errorfirst}}
[3cm]
{\resizebox{3cm}{!}{\begin{tikzpicture}

\def \a {5}       
\def \b {5}       

 \foreach \u in {0,1,...,\a}
    \foreach \v in {0,1,...,\b}
        \draw[very thin,white] (\u,\v,0.5) -- (\u,\v,0.5); 

\foreach \u in {0,4}
    \foreach \v in {0,1,...,4}
        \draw [fill= red!10,red!10] (\u,\v) rectangle (\u+1,\v+1);

\foreach \u in {0,4}
    \foreach \v in {0,1,...,4}
        \draw [fill= red!10,red!10] (\v,\u) rectangle (\v+1,\u+1);
    
\draw[style=help lines,dashed] (0,0) grid[step=1] (\a,\b);

\end{tikzpicture}}}
\hspace{0.1\textwidth}
\subcaptionbox{After two timestep, at $t=2\Delta t$\label{subfig:errorsecond}}
[3cm]
{\resizebox{3cm}{!}{\begin{tikzpicture}

\def \a {5}       
\def \b {5}       

 \foreach \u in {0,1,...,\a}
    \foreach \v in {0,1,...,\b}
        \draw[very thin,white] (\u,\v,0.5) -- (\u,\v,0.5); 

\foreach \u in {0,1,3,4}
    \foreach \v in {0,1,...,4}
        \draw [fill= red!20,red!20] (\u,\v) rectangle (\u+1,\v+1);

\foreach \u in {0,1,3,4}
    \foreach \v in {0,1,...,4}
        \draw [fill= red!20,red!20] (\v,\u) rectangle (\v+1,\u+1);
    
\draw[style=help lines,dashed] (0,0) grid[step=1] (\a,\b);

\end{tikzpicture}}}
\caption{Illustration of the propagation of Carlemann linearization error with timestep in a two-dimensional domain with uniform initial flow and non-periodic boundaries. Lattice cells with no fill have discrete exact discrete densities whereas ones with red fill have discrete densities with error}
\label{fig:errorpropagation}
\end{figure}
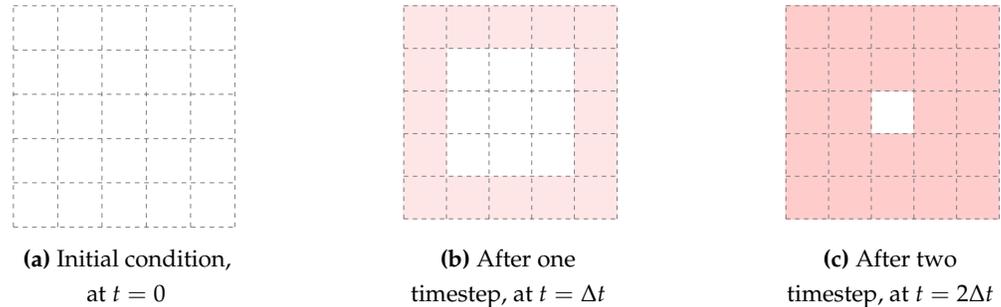

\section{Numerical Results}
\subsection{Logistic Equation}
\begin{figure}[!ht]
\includegraphics[width=\columnwidth]{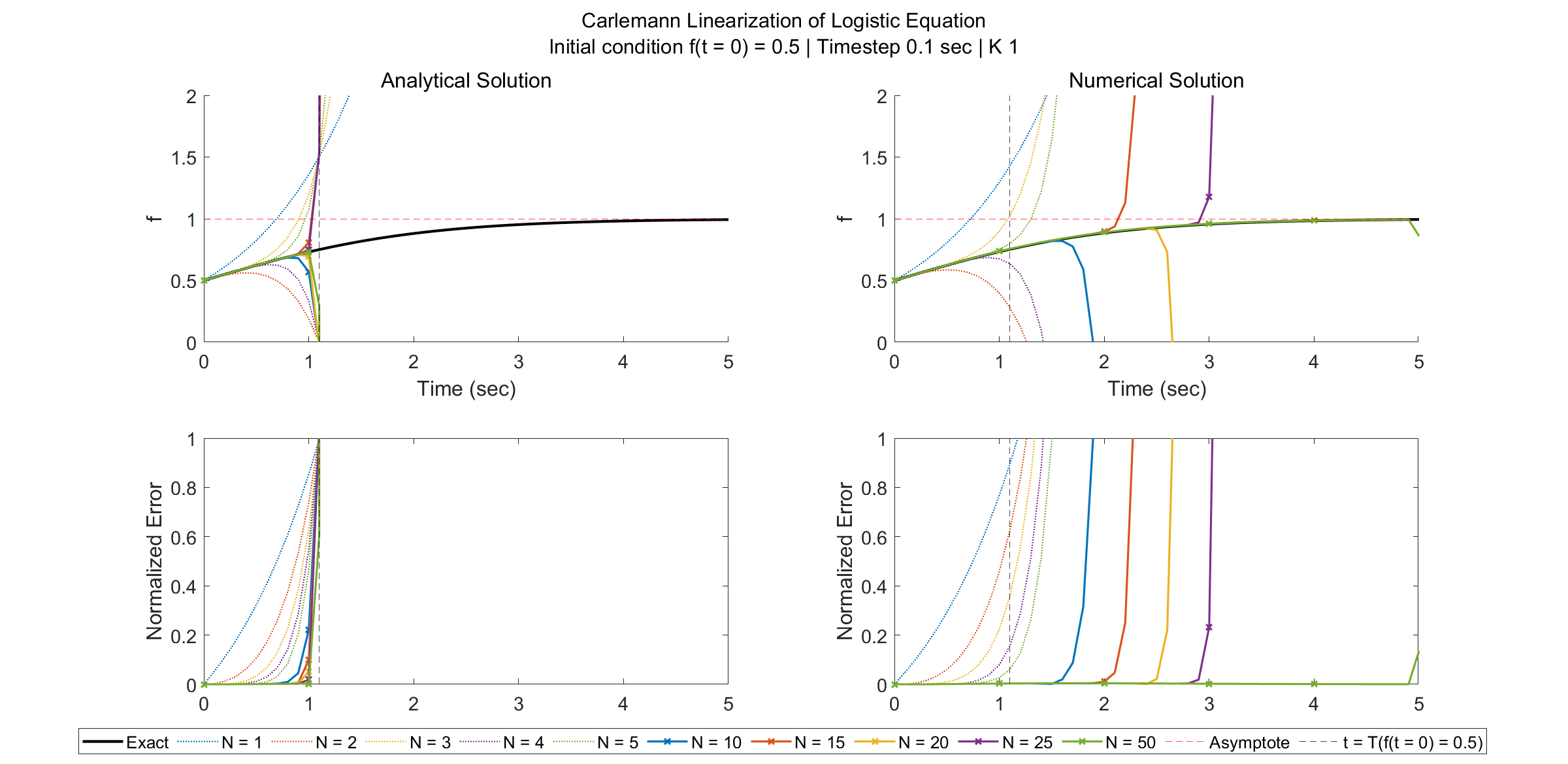}
\caption{The analytical (left) (analytical integration) and numerical (right) (discrete time-stepping) solutions of the Carlemann-linearized logistic equation are shown with their corresponding errors (bottom) as a function of time, varying initial conditions and Carlemann linearization orders. The predicted time of validity is shown as a vertical asymptote in each plot.}
\label{logFig}
\end{figure}

To demonstrate the utility of Carlemann linearization, we consider the logistic equation, as suggested by \cite{liuEfficientQuantumAlgorithm2020}.
\begin{equation}
\begin{matrix}
    \frac{df}{dt} = Kf(1-f) & \forall f \in [0,1]
    \end{matrix}
\end{equation}
We note that the $K$ factor appears for both first and second order terms, and, thus, cancels out in the calculation of $R$ which remains dependent on the initial conditions solely. In the results, we take $K=1$. We still see an abrupt cut-off in the utility of the linearization for the resulting analytical solution. This is explained by the fact that even though $R\leq 1$, the coefficient of the first-order term is definite positive, unity, thereby fulfilling neither $Re(\lambda_1)<1$ nor $\mu(F_1)<0$. According to \cite{foretsExplicitErrorBounds2017,liuEfficientQuantumAlgorithm2020}, this explains the blow-up in the error shown in the analytical solution. Namely, we restate upper-bounds for time $T$ using the power-series method of \cite{foretsExplicitErrorBounds2017}:
\begin{equation}
\label{Tend}
    T = \frac{1}{\norm{F_1}} \ln(1+\frac{\norm{F_1}}{\norm{f(t=0)}\norm{F_2}}) = \ln(1+\frac{1}{\norm{f(t=0)}})
\end{equation}
which predicts the time of validity for the linearization as shown in Table \ref{tab:4}, and with which the results agree, as the analytical solutions presented in Fig. \ref{logFig}.
\begin{table}[ht]
  \centering
    \caption{The maximum endtime validity for Carlemann linearization of the logistic equation as predicted by Eq. \ref{Tend}}
  \begin{tabular}{cc}
     $f(t=0)$ & $T$ \\
     \hline
    0 & $\infty$ \\
    0.1 & 2.40 \\
    0.2 & 1.79 \\
    0.3 & 1.47 \\
    0.4	& 1.25 \\ 
    0.5	& 1.10 \\
    0.6	& 0.98 \\ 
    0.7	& 0.89 \\
    0.8	& 0.81 \\
    0.9	& 0.75 \\
    1	& 0.69 \\
  \end{tabular}

  \label{tab:4}
\end{table}

However, for the numerical solution computed through time-discretization of the equation, we note the error shows slower evolution, giving a longer effective time-period to work with, as could be seen with the numerical solutions extending well-beyond their analytical counterparts before blowing up in Fig. \ref{logFig}. This is in line with the findings for the validity of the linearization of the Burger's equation with $R\approx40$, and an invitation for more applied work in the field.

\subsection{D1Q3}
\begin{figure}[ht]
\includegraphics[width=\columnwidth]{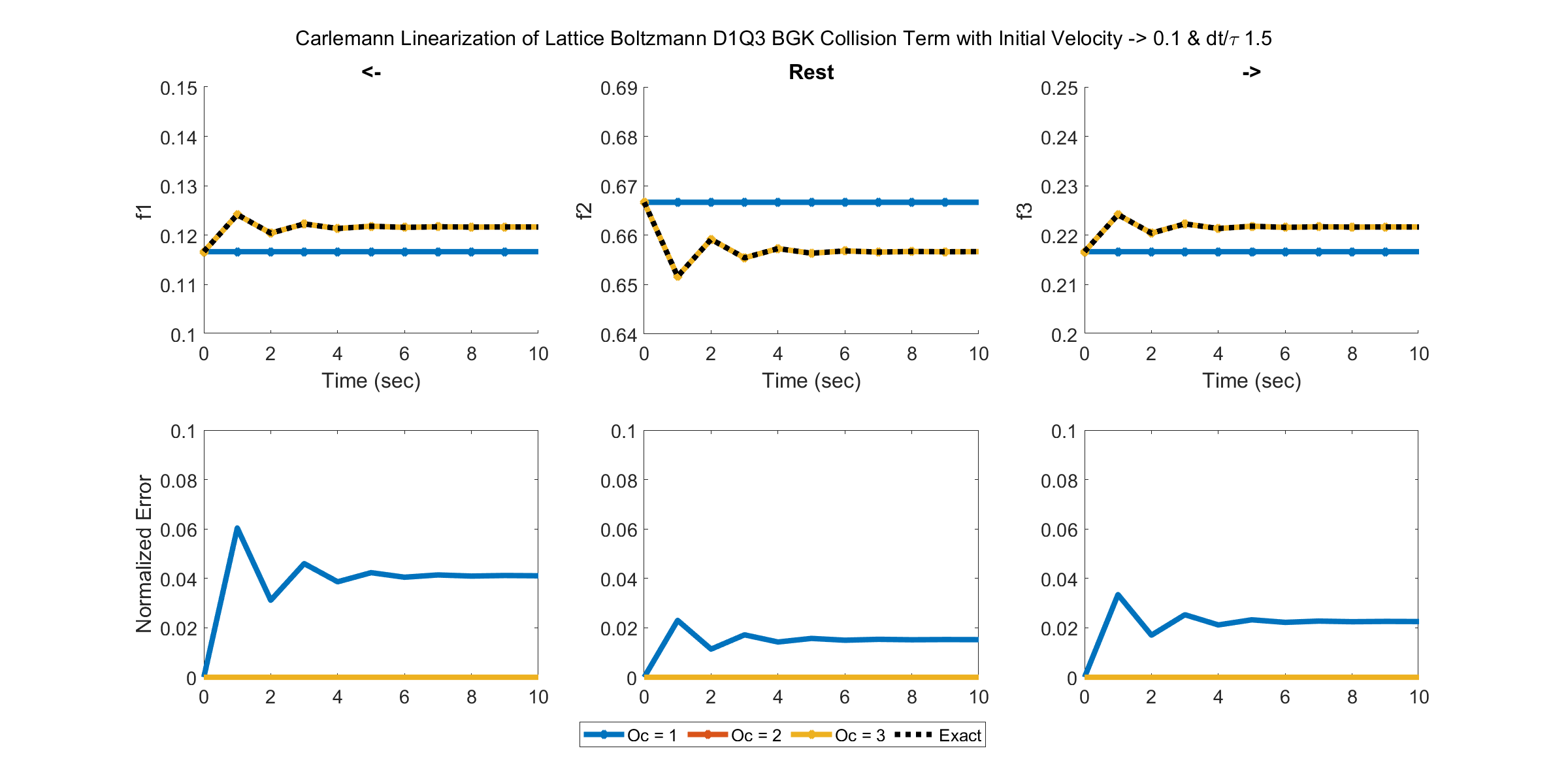}
\caption{The solution of the discrete densities of the fluid in D1Q3 for successive collisions is shown for the exact and Carlemann-linearized formulations as a function of time and Carlemann linearization order. The bottom figures show the normalized errors for each discrete density. Note that the solution is exact beyond the first linearization order.}
\label{d1q3res}
\end{figure}

We now concern ourselves with the results of linearizing the collision term of a D1Q3 lattice Boltzmann formulation. As mentioned in methodology section, exact streaming of the linearized system is only possible on a classical computer with the computation of the nonlinear terms, which defies the purpose of a linearization scheme. Therefore, we restrict ourselves to the collision step only. We note that in the absence of streaming,  we see in Fig. \ref{d1q3res} that the solution is exact for all orders of linearization starting from the second.

\section{Conclusion}
The classical algorithm suffers from a blowup in variable count and sacrifices the exactness of streaming. However, we have shown that the error of the classical Carlemann technique could be mitigated, even effaced, in specific applications. 
On the bright side, we have shown that, at least for the case explored in this paper, the error of the Carlemann linearization 
decreases exponentially with the order of the linearization. This bodes well for the development of a quantum LB algorithm
based on Carlemann linearization 
\cite{itani_quantum_2021}.


\newpage

\acknowledgments{The authors would like to acknowledge Dr. Antonio Mezzacapo for the long, fruitful discussions without which major findings of this paper would not have been possible.}

\funding{Sauro Succi acknowledges funding from the European Research Council under the European Union's Horizon 2020 Framework Programme (No. FP/2014-2020) ERC Grant Agreement No.739964 (COPMAT). Wael Itani acknowledges funding from New York University Tandon School of Engineering under the School of Engineering Fellowship 2021.}

\conflictsofinterest{The authors declare no conflict of interest.} 

\dataavailability{The MATLAB codes used in this study are openly available in GitHub at \url{https://github.com/waelitani/carlemann-linearization-lattice-boltzmann}}

\newpage
\printnomenclature

\newpage
\reftitle{References}

\bibliography{references.bib}
\end{paracol}
\end{document}